# Soft Matter

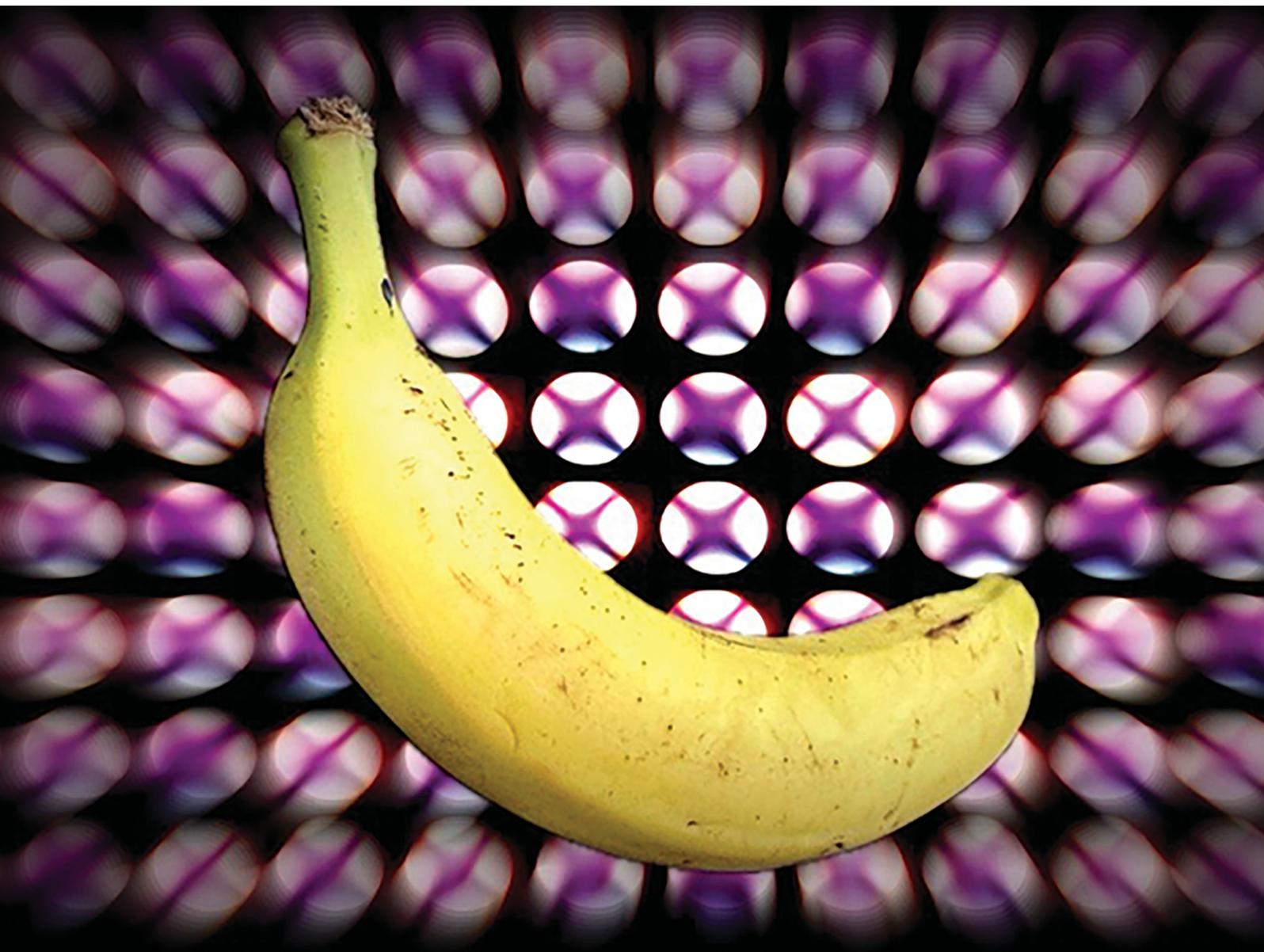

ROYAL SOCIETY OF CHEMISTRY

**PAPER**
Dorota Węgłowska *et al.*
Banana DNA derivatives as homeotropic alignment layers in optical devices

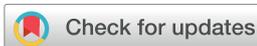

# Banana DNA derivatives as homeotropic alignment layers in optical devices

Rafał Węgłowski, Anna Spadło and Dorota Węgłowska  *

In this study, deoxyribonucleic acid (DNA) from bananas was extracted and functionalized and used for the first time as a homeotropic alignment layer for liquid crystals (LCs). Our research was aimed at extracting and investigating DNA from bananas *via* the synthesis and study of DNA complexes with various surfactants to examine the usefulness of such a complex as an alignment layer in electro-optical transducers. We proposed a simple and eco-friendly synthesis of the DNA complexes isolated from bananas with surfactants, so we transformed the DNA isolated from bananas into a functionalized alignment layer. A biopolymer alignment layer like deoxyribonucleic acid (DNA) from a banana complexed with a cationic surfactant is an excellent alternative to a commonly used but toxic polyimide alignment layer. DNA-based materials are promising for photonic applications and biosensors because of their excellent optical and physical properties, biodegradability, and low production cost. The novelty of the research lies in the potential use of these materials as biodegradable biopolymer alignment layers for optical devices instead of conventional polymers, which are usually harmful for the environment.

## 1. Introduction

The growing population and demand for processed food have significantly contributed to the food waste (FW) phenomenon. FW is a global problem, accounting for the waste of one-third of the total food produced for human consumption (1.3 billion tonnes per year).[1] Improper management and improper food waste disposal impose an additional burden on the environment and lead to negative consequences for living things, along with a severe economic crisis.[2,3] Fruit and vegetable waste accounts for a significant amount of waste generated (42%) in various proportions of food product waste.[4] Global banana (*Musa paradisiaca*, family Musaceae) production was approximately 119 million tonnes in 2020.[5] About 30–40% of whole bunches of bananas generate this waste, which means that around 42 million tons of fruit waste is generated annually worldwide. Although fresh bananas can be eaten, they are also processed into various products. Green bananas can be used for making banana flour, while ripe bananas can be processed into syrup, juices, and dried bananas.[6] FW has great potential to meet the growing demand for biopolymers and replace petroleum-based synthetic polymers.[7] Synthetic plastic polymers are very stable and versatile, exhibit excellent performance under adverse situations, and are inexpensive, which explains their widespread use worldwide.[8] Petroleum-derived plastic waste is non-biodegradable and harmful to all life and nature. As bio-feedstock for synthesizing biopolymers, FW has gained popularity as a tangible and sustainable alternative to the petroleum-based economy due to its non-toxic, biocompatible, and biodegradable nature.[9] Biopolymers such as polyhydroxyalkanoates (PHA), polyhydroxybutyrates (PHB), polysaccharides (PS), polynucleotides, polylactic acid (PLA), and polypeptides are recovered from FW as a substitute for traditional petroleum-derived plastics.[10] However, very little information has been documented regarding using food-based products in the context of optical devices like transducers or biosensor applications and waste reduction for a clean environment. This clearly shows the research gap in producing biopolymers from various food wastes and their implications. Detailed studies were conducted on producing banana biopolymers to overcome the lack of knowledge. Then, an attempt was made to use modified banana DNA as a liquid crystal alignment layer in LC optical devices.

Nematic liquid crystals (NLCs) are excellent materials for optical transducers or high sensitivity, low-cost optical biosensors.[11,12] A NLC with negative dielectric anisotropy is used in the homeotropic alignment of molecules, where the molecules are perpendicular to the surface. Therefore, the LC molecules reorient themselves perpendicular to the applied electric field. In LC biosensors, the change in the orientation of the LC optical axis is caused by placing the analyte (*e.g.*, a biomolecule) on one of the biosensor substrates. This, in turn, changes the transmission of light passing through the liquid crystal cell. Such biosensors do not need an external electric field.[13,14]

*Faculty of Advanced Technologies and Chemistry, Military University of Technology, 2 Kaliskiego Str., 00-908, Warsaw 49, Poland.*
*E-mail: dorota.weglowska@wat.edu.pl*

One of the critical elements of such a liquid crystalline transducer/biosensor is an alignment layer that properly aligns the liquid crystal material's optical axis. Petrochemical industry products produce traditional polymers as the LC alignment layer, and the properties of synthetic polymers do not always allow the used material to be recycled. Synthetic polymers such as polyimide (PI), which are used as homeotropic alignment layers, are long-life materials exhibiting high-temperature stability (up to 350 °C) but are not produced using environmentally friendly methods.[15–17] The organic solvents used in production and under high thermal conditions with high energy process consumption are often out of green chemistry criteria,[18] and monomers, particularly diamines and their intermediates, are highly toxic and carcinogenic. Therefore, it is crucial to use a bio-renewable alignment layer of LCs in a non-toxic synthetic way.

The excellent "candidates" for substituting PI-based layers are biopolymers, such as deoxyribonucleic acid (DNA) derivatives. Such specific biopolymers are under extensive research for many applications, especially outside the biological field of interest. Some of the research is concentrated on the form of DNA–cationic surfactant complexes. The structure consists of DNA and different surfactants attached, through electrostatic interactions, to a negatively charged phosphate residue. DNA complexes are essential for numerous photonics and optics applications.[19–21] Using a biodegradable DNA biopolymer as an element of the liquid crystal alignment layer is advantageous because used sensors will be disposed faster. DNA complexes with surfactants, when vertically oriented in liquid crystals, exhibit desirable application properties, such as high thermal stability and excellent light transmission in the visible and near-infrared ranges. A DNA complex with a surfactant forms quickly at room temperature in water, which is an additional advantage (no need to dispose of used organic solvents).

## 2. Experimental

### 2.1 Isolation and identification of DNA from banana

Bananas collected from the local supermarket (Warsaw, Poland) were used as feedstock for complexes of DNA-based alignment layers. The first stage of the experimental work was the separation/attempt of DNA isolation of the banana. Peeled bananas (280 g) were cut and placed in a 250 ml beaker. They were then crushed using a blender. Then, sodium chloride (8.5 g), hot water (100 ml), and dishwashing soap (8 g) were added and gently stirred. The resulting mixture was heated (65 °C) in a water bath for 15 minutes and then cooled for 15 minutes in ice and filtered using a Büchner funnel under reduced pressure. The sediment on the filter was removed and thrown away. The filtrate was transferred to the beakers and then, cold isopropanol (volume 1:1) was added by carefully pouring down the sides of the beakers, causing the DNA to precipitate. The precipitate was in the form of gelatinous, white, long threads with air bubbles between them. Finally, the DNA was washed several times with cold isopropanol and allowed to dry at room temperature until constant weight.

### 2.2 Materials

Two cationic surfactants such as dimethyldioctadecylammonium chloride (DODA; MW = 586.5 g mol$^{-1}$, with a purity of >97%; Sigma-Aldrich, Co.) and benzyldimethylhexa-decylammonium chloride (CBDHA; MW = 396.1, with a purity of 99%; Sigma-Aldrich, Co.) were chosen to modify DNA sodium salt. Commercial polyimide SE-1211 (Nissan Chemical Industries, Ltd.) was used as a reference alignment layer. To check the properties of alignment layers of substrates and complexes, the 1832 nematic mixture ($\Delta n$ = 0.23 at 589 nm and 20 °C; and negative dielectric anisotropy $\Delta \varepsilon$ = −11.1 at 20 °C; prepared at the Military University of Technology) consists of 1-alkoxy-4-[(4-alkyl phenyl)ethynyl]benzene and 2,3-difluoro-1-alkoxy-4-[(4-alkyl phenyl)ethynyl]benzene with different alkoxy- and alkyl chain lengths was used.

### 2.3 Synthesis of biopolymers

DNA forms complexes with cationic surfactants using electrostatic interactions without a formal coordination bond, requiring oppositely charged ions. DNA sodium salt in an aqueous solution has a negative charge that quickly creates a complex with a cationic surfactant in a two-step process. First, the surfactant molecule binds on electrostatic interactions to the phosphate residues of the DNA structure. Then, the surfactant groups are positioned along the DNA structure due to strong hydrophobic interactions between the surfactant groups. DNA–surfactant complexes were synthesized based on the procedure described in ref. 22. The DNA sodium salt and the selected surfactant were dissolved in equal amounts (by weight) in deionized water. Then, the surfactant solution was added to the DNA solution drop-by-drop using a magnetic stirrer. After filtration, the DNA–surfactant complexes were dried at room temperature. The schematic reaction of obtaining the DNA with selected surfactants (DODA and CBDHA) is presented in Fig. 1.

### 2.4 UV-vis measurements

UV-vis measurements determined the quality of the extracted DNA. The absorption spectra of DNA isolated from bananas (25 μM; in water), surfactants (25 μM; in water), and DNA complexes with surfactants (25 μM; in butanol) were investigated (180–1000 nm), using a UV-vis-NIR 3600 Shimadzu spectrophotometer (Nakagyoku, Kyoto, Japan). The absorption measurements were recorded in the range of 205–325 nm. A blank sample was recorded for an empty quartz cuvette (1 cm thickness). All samples were normalized to the maximum value.

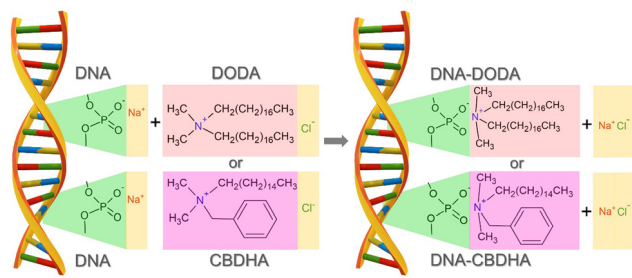

Fig. 1 The schematic reaction for obtaining DNA with selected surfactant complexes.

### 2.5 IR measurements

Infrared measurements of DNA isolated from bananas, surfactants, and DNA complexes with surfactants (region 2–6 μm) using reflective mode from crystals using the ATR technique with a diamond crystal were carried out, and a Nicolet iS10 Thermo Scientific spectrometer was used. The spectra were recorded using a diamond crystal within a wavenumber range of 400–4000 cm$^{-1}$ with a resolution of 2 cm$^{-1}$ from 24 scans. Before each sample measurement, the background measurement was recorded.

### 2.6 Polarizing optical microscopic (POM) study

To check the alignment properties, the liquid crystalline textures of different cells, prepared with the commercial polyalignment layer SE-1211 and with DNA, surfactants, and DNA–surfactant complexes were observed by polarized optical microscopy (POM, OLYMPUS Optical Co., Ltd, Models BHSP-2, BX-51). The liquid crystal cells with alignment layers were placed between two crossed polarizers. Observations of the liquid crystal alignment were carried out in both orthoscopic and conoscopic modes. Optical conoscopy was performed by inserting an Amici-Bertrand lens positioned in the optical system of the polarized light microscope.

### 2.7 Preparation of a liquid crystal cell with an alignment layer

The powder of DNA with surfactant complexes was dissolved in butanol (3 wt%) and then spin-coated onto a glass plate covered with an indium tin oxide (ITO) conductive layer. The coating was heated at 80 °C for 1 h to evaporate the solvent. Two substrates with an alignment layer were glued together to obtain a measuring cell. The glue was mixed with spacers of 5 μm thick to ensure the appropriate cell thickness. Then, the cell was filled with liquid crystals by a capillary action. This procedure was the same for the DNA with surfactants and commercial polyimide layers.

Typical liquid crystalline transducers consist of two glass substrates with transparent indium tin oxide (ITO) to allow the application of an electric voltage (field) across the cell. The glass cell is filled with an optical anisotropic birefringence liquid crystal and placed between two crossed polarizers. LC birefringence is defined as the difference between an ordinary refractive index ($n_o$) and an extraordinary refractive index ($n_e$) (Fig. 2.). The response of the nematic liquid optical axis to the applied field depends on the dielectric anisotropy. A liquid crystal with negative dielectric anisotropy is used in the homeotropic alignment of molecules, where the molecules are perpendicular to the surface. Therefore, the LC molecules reorient themselves perpendicular to the applied electric field.

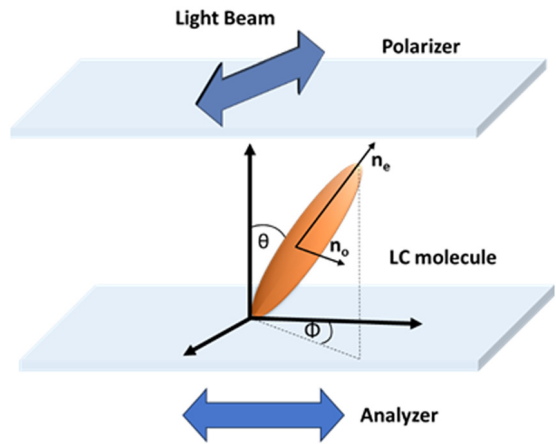

Fig. 2 The schematic drawing of the LC cell with a polarization-dependent refractive index. $n_o$ is the ordinary refractive index, $n_e$ is the extraordinary refractive index, $\Phi$ is the azimuthal angle of the LCs against the surface, $\theta$ is the angle between the director of the liquid crystal and the propagation direction of light.

The phase retardation of light passing through such a cell is described using the following equation:

$$\delta = \frac{2\pi d \Delta n_{\text{eff}}}{\lambda} \quad (1)$$

where $\delta$ is phase retardation, $d$ is the thickness of the liquid crystal layer whose arrangement has been disturbed, $\lambda$ is the wavelength of incident light, $\Delta n_{\text{eff}}$ is the effective birefringence of the liquid crystal which can be described as follows:

$$\Delta n_{\text{eff}} = \frac{n_e n_o}{\sqrt{(n_e \sin\theta)^2 + (n_e \cos\theta)^2}} \quad (2)$$

where $\theta$ is the angle between the director of the liquid crystal and the propagation direction of light.

The applied electric field does not affect the refractive index of the ordinary ray passing through the cell. However, the magnitude of the refractive index of the extraordinary ray increases with voltage due to the dielectric coupling of the nematic director with the field if the liquid crystal mixture exhibits negative dielectric anisotropy. Thus, the effective birefringence increases with the applied electric field strength.[13]

The transmission of light traversing the cell is described by the following equation:

$$T = \frac{I}{I_0} = \sin^2 2\phi \sin^2\left(\frac{\delta}{2}\right) \quad (3)$$

where $I$ is the intensity of light coming from the LC cell, $I_0$ is the intensity of light illuminating the LC cell, and $\Phi$ is the azimuthal angle of the LCs against the surface.

With an applied voltage, the vertically aligned LC molecules are switched in a direction parallel to substrates and under ideal conditions of $\phi = \pi/4$ and $\delta = \pi$ to maximize the light transmittance, that is, $T = 1$.

For zero voltage applied to the cell, the polarization of the linearly polarized incident light is perpendicular to the long axes of LC directors, and then the light experiences only the

Paper    Soft Matter

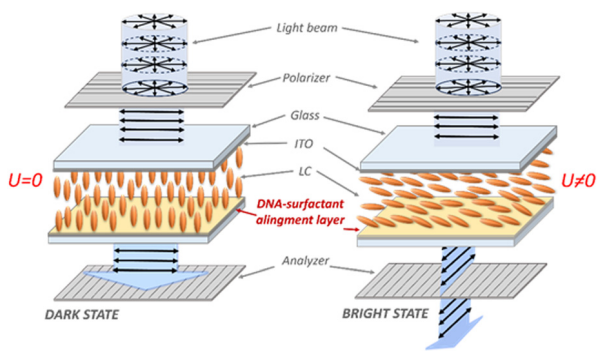

Fig. 3 The schematic drawing of the driving of an electro-optical liquid crystal transducer.

refractive index of $n_o$. Consequently, the polarization of the light passing through the cell does not change. There is the dark state. Applying a voltage above the threshold voltage $V_{th}$ causes the nematic director to tilt away from the normal and modifies the phase retardation and polarization of the outgoing light. Therefore, a proportion of the polarised light traverses the analyzer, and the bright state is obtained (Fig. 3).

### 2.8 Electro-optical measurements

An optical polarizing microscope Olympus BX51 in transmission mode was used. LC cells were placed between two crossed polarizers. White light from a halogen lamp was then passed through this configuration. An AC electrical field (0–10 V, 1 kHz, square wave) was applied to the cells using a function generator (HP 33120A). A linear optical detector recorded light intensity and monitored using an oscilloscope (HP 346013) connected to a computer. The data were acquired using the AgilentVee Software.[23] The measurements were performed at room temperature.

### 2.9 Ionic conductivity measurements

Our work examined the influence of DNA–complexes alignment layers on ionic contamination in a liquid crystal. For this purpose, the electrical conductivity of the liquid crystal was measured in the cells with DNA–DODA, DNA–CBDHA, and SE1211 DNA layers. The dielectric spectrum was acquired using a Hewlett-Packard 4192A impedance analyzer in a parallel equivalent circuit.[24] The G conductance measurements were conducted over a broad frequency range of 10 Hz to 10 MHz. A measuring electric field of 0.1 V was applied across the measuring 5 μm thickness cells.

## 3. Results and discussion

### 3.1 UV-vis measurements

The UV-vis absorption spectra of DNA isolated from bananas (water solution), used surfactants (water solution), and complexes with surfactants (butanol solution) are shown in Fig. 4.

The most common calculation of the purity of the isolated DNA is the ratio of the absorbance at 260 nm divided by the reading at 280 nm. Good quality DNA exhibits an A260/A280 ratio of 1.7–2.0. The absorbance ratio (A260/A280) of DNA isolated from bananas is 1.21, which indicates that DNA isolated from bananas is contaminated. Possible contaminations that could be in the isolated DNA are RNA and aromatic amino acids present in the proteins, which exhibit $\lambda_{max}$ at 260 nm, the presence of peptide bonds and aromatic amino acids in proteins with the absorption of 210–230 nm (peptide bond) and 260–280 nm (aromatic amino acids). The $\lambda_{max}$ below 230 nm is the result of the presence of peptide bonds in proteins and, to some extent, Trp, Tyr, Phe, Cys–Cys, His, Met and Arg residues. Strong absorbance around 230 nm may indicate the presence of organic compounds or chaotropic salts. The ratio of 260 nm to 230 nm can help to assess the level of salt transfer in purified DNA. A260/A230 is best if it is higher than 1.5. The absorbance ratio (A260/A230) of DNA isolated from bananas is 0.42.

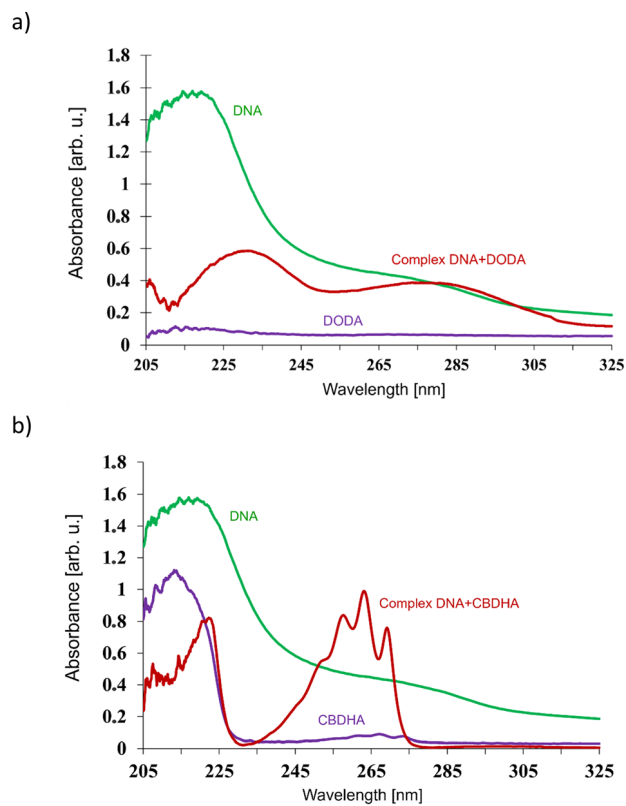

Fig. 4 UV-vis absorption spectra of 25 μM DNA in water, 25 μM surfactants, and 25 μM DNA complexes with surfactants in butanol: (a) surfactant/complex DODA and (b) surfactant/complex CBDHA.

The observed absorption below 225 nm results from the absorption of the phosphate groups and the deoxyribose moieties[25] occurring both in DNA and complex DNA with surfactants. Still, this $\lambda_{max}$ position of absorption spectra of 260 nm is observed for the complexes with both surfactants, meaning that the reaction between the DNA and the surfactant is selective and allows DNA purification. This absorption comes from electronic transitions of aromatic bases of DNA.[26]

There is a presence of proteins or other contaminants with an absorbance close to 280 nm. Contaminants of the DNA

isolated from the banana during the reaction remain in the water. The surfactant CBDHA, in contrast to the surfactant DODA, exhibits aromatic ring absorption below 225 nm.

### 3.2 IR measurements

Infrared absorption spectroscopy was used to confirm the proper chemical composition of compounds used to prepare DNA–surfactant complexes and the obtained complexes.

The comparison of the IR spectra of DNA isolated from bananas, DODA, CBDHA, and DNA–DODA and DNA–CBDHA complexes is presented in Fig. 5. The spectra presented in Fig. 5 are average spectra from five random points on the surface of each sample.

The band that is most sensitive to structural changes during complex formation is the band associated with $\nu_{as}$ ($PO_2^-$) vibrations.[27] A shift of this band towards higher wave numbers (above 1240 cm$^{-1}$) is observed for both DNA with surfactant complexes, which confirms their formation. The shape of the band in the spectral range of 1080–1120 cm$^{-1}$, in which the $\nu_s$ ($PO_2^-$) vibrations are present, is also modified.

The IR spectrum of DNA extracted from bananas is similar to the IR spectrum of commercial salmon DNA.[27] Still, the IR absorption bands appearing, among others, in the areas of 1225–1220 cm$^{-1}$ ($PO_2^-$, the main marker of the β form) and 1053 cm$^{-1}$ (C-O in deoxyribose) are much more intense for commercial DNA than for the DNA isolated from bananas, which indicates the presence of impurities. Compared to the DNA isolated from the banana spectrum, the following absorption bands appear in both surfactants and DNA complexes with surfactants, 2920 and 2850 cm$^{-1}$. These bands correspond to symmetric and asymmetric stretching C–H vibrations of the $-CH_2$ and $-CH_3$ groups. The peak observed at 1472 cm$^{-1}$ in DODA, CBDHA, and both complex spectrum corresponds to symmetric stretching C–H vibrations of $-CH_2$ and $-CH_3$ groups. The peaks observed at 729 and 700 cm$^{-1}$ in CBDHA, and the DNA–CBDHA complex spectrum correspond to stretching C–C skeletal vibrations in the phenyl ring. The broad bands at around 3400 cm$^{-1}$ (O–H stretching vibration) in DODA, CBDHA, and the DNA–CBDHA complex spectrum are evidence of the presence of water left. These peaks in both the surfactant and complex spectra confirm that the surfactants are bound to the DNA in the complex sample.

### 3.3 LC orientation behavior of the LC cell

Fig. 6 shows images of LC cells placed between crossed polarizers, made with DNA, surfactants, DNA–surfactants and SE-1211 layers to observe the orientation behavior of the LC molecules. The LC cells with DNA films show the irregular arrangement of the optical axis of the liquid crystal layer. The same effect can be observed for LC cells containing DODA and CBDHA surfactant layers. However, DNA–surfactant complexes show excellent results in homeotropic LC ordering. Dark colors of the samples over the entire surface are related to the vertical alignment of the LC in the cell. A similar situation can be observed in a liquid crystal cell with a commercial SE-1211 alignment layer. This means that, at first glance, DNA–surfactant complexes are a good alternative to commercial alignment layers. Moreover, the vertical alignment of the liquid crystal in cells containing DNA–surfactant complexes did not change after several months.

### 3.4 Polarizing optical microscopic (POM) study

The more precise measurements of LC aligning behaviors of the LC cells made from DNA, surfactants, and DNA–complex layers were also examined by observing orthoscopic and conoscopic POM images, as shown in Fig. 7. Without an external electric field we can see that the uniform black color of LC textures for DNA–DODA and DNA–CBDHA cell samples confirms the homeotropic alignment of LC molecules throughout the cell under crossed polarizers. When LC molecules are anchored homeotropically on alignment layers, the polarization state of incident light through the polarizer remains unchanged. Consequently, the analyzer blocks the light, and a dark state can be observed. The isolated DNA, surfactant DODA, and CBDHA alignment layers do not show the ability to orient the liquid crystal molecules homeotropically. In this case, many surface defects cause changes in the anchoring of LC molecules, which cause microscopic, random disturbances in their ordering. The same effect was observed for a cell without orientation layers, containing only the bare ITO layer. Such a cell did not show homeotropic liquid crystal ordering even after several weeks (Fig. 7a).

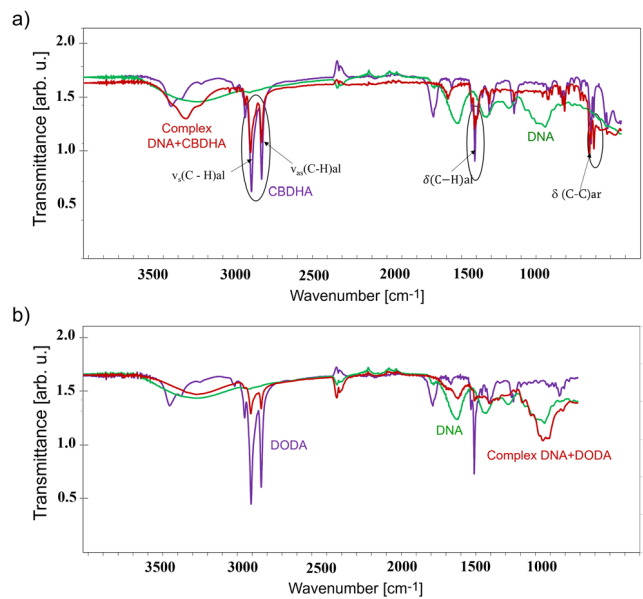

Fig. 5 IR spectra of DNA, surfactants, and DNA complexes with surfactants: (a) surfactant/complex DODA and (b) surfactant/complex CBDHA.

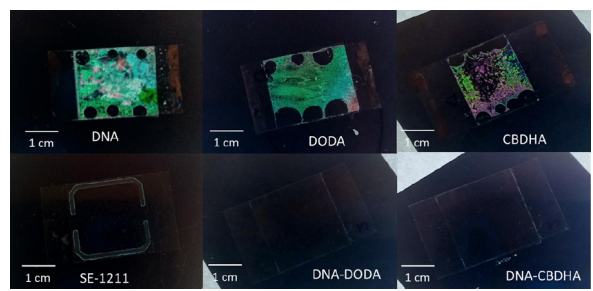

Fig. 6 Polarizing optical microscopy images of the LC cell placed with different alignment layers between the crossed polarizers.

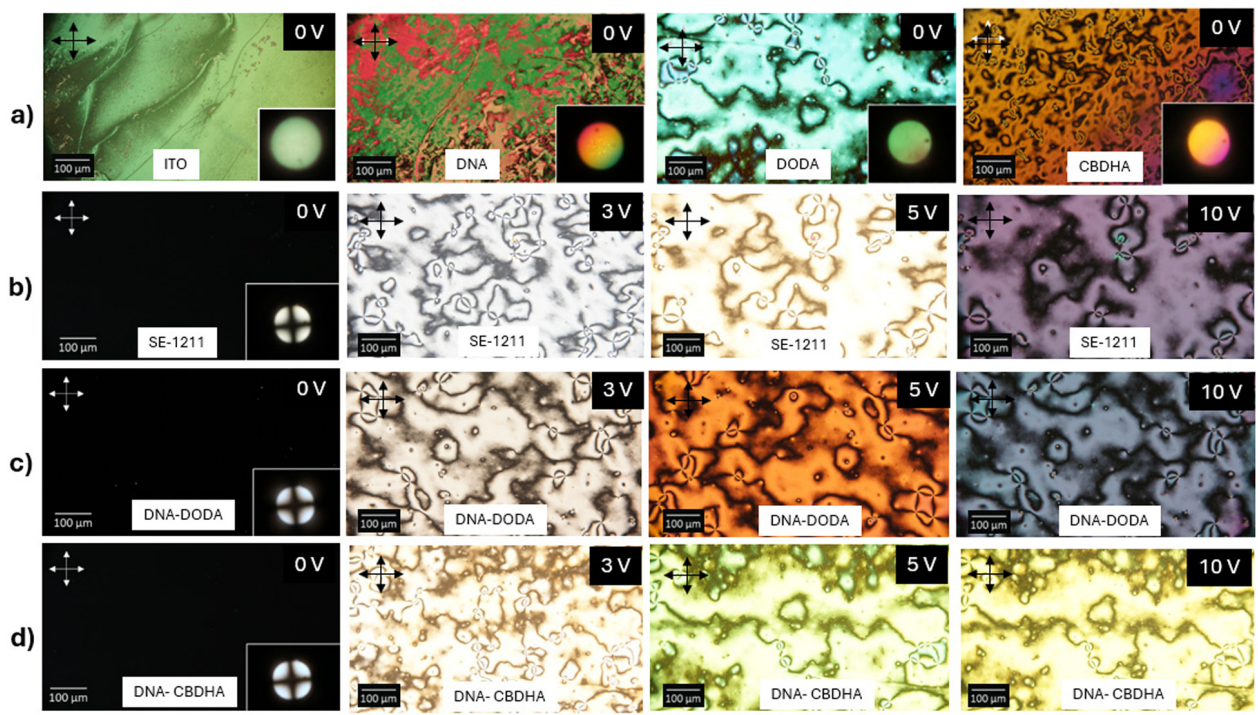

**Fig. 7** Orthoscopic and conoscopic optical polarizing microscopy images of the LC cell placed with different alignment layers between crossed polarizers: (a) the bare ITO layer, isolated DNA, surfactant DODA, and CBDHA alignment layers without voltages, (b) the SE-1211 alignment layer at different voltages, (c) the DNA–DODA alignment layer at different voltages, (d) the DNA–CBDHA alignment layer at different voltages.

To further analyze the quality of the homeotropic ordering of the liquid crystal, conoscopic observations were used. Optical conoscopy has traditionally been used to investigate the orientation of optic axes in liquid crystal materials.[28] When the optic axis of the nematic liquid crystal is parallel to the light path, it forms a Maltese cross in the center, as is shown for LC cells with no voltage applied (Fig. 7b–d), respectively.

The Maltese cross pattern is a consequence of the interplay between the polarized light, the birefringence of the nematic liquid crystals, and the symmetrical alignment of the LC molecules. The LC molecules are uniformly aligned vertically, causing a symmetrical distribution of the interference pattern.

The cross's arms correspond to the directions where the light components interfere destructively (resulting in dark areas), while the regions in between show constructive interference (resulting in bright areas). This symmetrical interference pattern creates the characteristic Maltese cross shape.[29] The Maltese cross patterns in the conoscopic POM images of the LC cells indicate no discernible differences in the LC orientation of SE-1211, DNA–DODA and DNA–CBDHA ordering layers. The excellent homeotropic alignment of several-micrometer LC cells allows for a DNA-complex orientation layer in optical transducers and biosensors. When the optical axis of the liquid crystal molecules is arranged non-uniformly, the planar Maltese cross is not visible in Fig. 7a.

The optical texture of LCs in the SE-1211, DNA–DODA and DNA–CBDHA coated LC cells under a polarising optical microscope for the example applied voltage has been observed as shown in Fig. 7b–d, respectively. At zero voltage, the textures of prepared LC cells have shown an excellent dark state under crossed polarizers because the light passing through the LC layer does not experience birefringence. The light remains polarized in the same direction and is blocked by the analyzer. Upon applying the external voltage, the dark state was observed at approximately up to 2.0 V for all samples. The further increase in voltage exerts torque on the LC molecules and causes them to realign according to the field's direction. As a result, the birefringence of the LC increases, leading to an increase in brightness textures due to the greater amount of light transmitted. Moreover, the change in birefringence modifies the interference pattern of the polarized light. The interference pattern determines the color observed under a polarizing microscope. When the LC molecules reorient due to the electric field, the change in birefringence leads to a shift in the interference colors. As a result, the color observed through the polarizing microscope changes (Fig. 7b and c). When an electric field is present due to the unrubbed alignment surface, the LC molecules collapse onto the surface with a random azimuthal direction. Subsequently, the LC molecules reorient themselves to minimise free energy by altering the positions of disclination lines and point defects, which can be seen as dark lines.

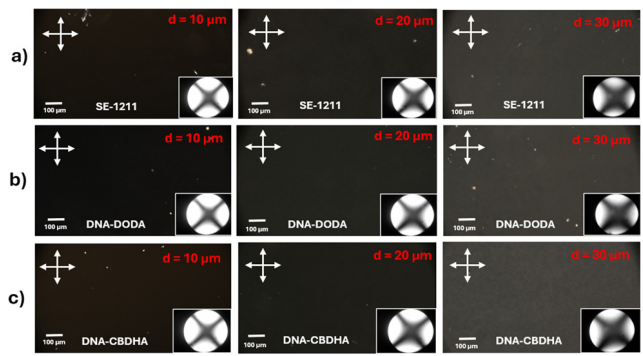

Fig. 8 Orthoscopic and conoscopic optical polarizing microscopy images of the LC cells with different $d$ thicknesses: (a) the SE-1211 alignment layer, (b) DNA–DODA alignment layers, (c) DNA–CBDHA alignment layers.

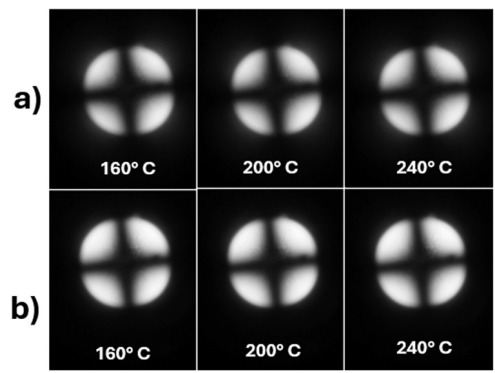

Fig. 9 Conoscopic optical polarizing microscopy images of the LC cells after thermal treatment at 160 °C, 200 °C, and 240 °C for 20 minutes: (a) the DNA–DODA alignment layer and (b) the DNA–CBDHA alignment layer.

The influence of the thickness $d$ of liquid crystal cells on the quality of the liquid crystal alignment was also examined. According to eqn (1), as the thickness cell increases, the phase shift increases proportionally, leading to larger regions where significant phase differences occur. Larger phase shifting means more pronounced constructive and destructive interference. The arms of the Maltese cross represent regions of destructive interference, and with the increased phase difference, these regions become more extended, resulting in thicker arms. This enlarges the overall size of the Maltese cross (Fig. 8). Moreover, we can notice an increase in the amount of light passing through the 30 μm thick layer. A change in the anchoring conditions and arrangement of the optical axes of liquid crystal molecules in the liquid crystal layer causes this slight light leakage.

Further research determined the influence of high temperatures on the alignment properties of the DNA–DODA and DNA–CBDHA layers. For this purpose, 5 μm thickness empty cells without liquid crystals were heated at 160 °C, 200 °C, and 240 °C for 20 minutes. After the heating process, the cells were filled with a nematic liquid crystal and its orientation was observed using a polarizing optical microscope. As shown in Fig. 9, no differences in the homeotropic LC orientation on the DNA–DODA and DNA–CBDHA layers were observed through the Maltese cross pattern in the conoscopic POM images. This indicates that the vertical LC orientation was maintained in this kind of LC cells even after exposure alignment layers to high temperatures.

### 3.5 Electro-optical measurements

To investigate the possibility of using an LC cell with a biological orientation layer as an optical transducer, electro-optical measurements were carried out. The voltage-dependent transmittance properties of LC cells with DNA–DODA, DNA–CBDHA, and the reference (commercial SE-1211) alignment layers are shown in Fig. 10.

Initially, every LC cell transmission intensity remains almost constant in the 0–1.8 V voltage range. There is no effective birefringence ($\Delta n_{\text{eff}} = 0$) since $n_e = n_o$. LC molecules are aligned with the electric field with a further increased applied voltage.

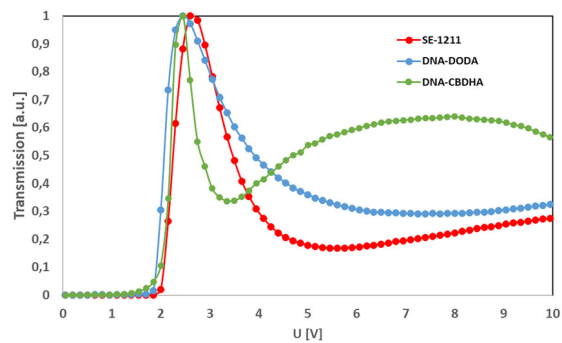

Fig. 10 The voltage-dependent transmittance of LC cells with three different homeotropic alignment layers.

This induces an effective birefringence, which is proportional to the magnitude of the applied field up to the maximum value of $\Delta n$. As a result, the transmittance is gradually increased because phase retardation occurs due to the different propagation speeds of extraordinary and ordinary rays in the liquid crystal medium. The threshold voltage of each LC cell is close to 2 V, and the transmission becomes almost saturated at close to 2.45 V for both DNA complexes and 2.75 V for commercial PI. The light transmission characteristics are a function of $\sin^2(\delta/2)$, where $\delta = 2\pi\Delta n/\lambda$ is the phase retardation. Therefore, for a cell with $d\Delta n > \lambda/2$, the transmission decreases with the increasing voltage value. The shape of the transmission curve for the cell with a DNA–CBDHA complex alignment layer is slightly different from those with a DNA–DODA complex alignment layer and a SE-1211 polyimide complex alignment layer. This is probably due to slightly different anchoring conditions of the LC molecules on the double aliphatic chain of the DODA surfactant. Detailed studies of the anchoring energy of LC molecules on biological orientation layers will be performed in the following research stage.

### 3.6 Ionic conductivity

In the experiment, the real part of conductivity $\sigma_{\text{EXP}}$ is defined as follows:

$$\sigma_{\text{EXP}}(f) = G\frac{d}{S} \qquad (4)$$

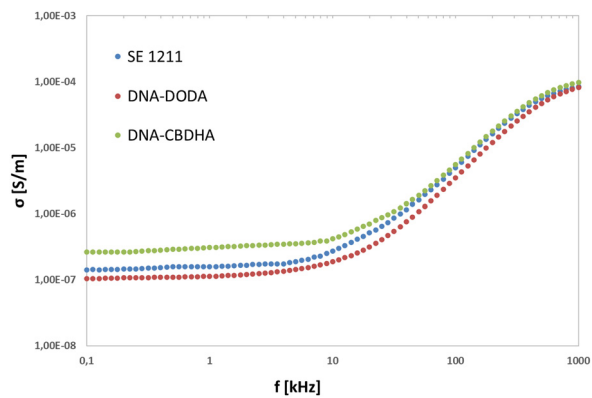

Fig. 11 Experimental dependence of electrical conductivity $\sigma$ upon frequency $f$ of the measurement field in SE-1211, DNA–DODA and DNA–CBDHA cells. The data were obtained in the nematic phase at 25 °C.

Here, $G$ represents the measured conductance using an impedance analyzer, a crucial tool in our experimental setup. The variables $d$ and $S$ denote the thickness and the active ITO electrode area of the cell, respectively. Fig. 10 presents the frequency dependence of $\sigma_{EXP}$ in the nematic phase at room temperature. The experimental data aligns with the theoretical model proposed by A. K. Jonscher[30] for a real part of conductivity:

$$\sigma_{EXP}(f) = \sigma_{DC} + a2\pi f^m \quad (5)$$

where $\sigma_{DC}$ corresponds to a plateau value of $\sigma_{EXP}$ within a certain frequency range, $a = \varepsilon' \varepsilon_0$ denotes a constant which depends on temperature and $m$ is a frequency exponent. The frequency dependence of $\sigma_{EXP}$ allows us to extract the value of the static (DC) ionic conductivity.[31]

The experimental data on the frequency dependence of electrical conductivity for cells with SE-1211, DODA CBDHA, and DNA–DODA alignment layers are shown in Fig. 11. In the frequency range from 0 and 1 to 10 kHZ, $\sigma$ is practically constant. For frequencies higher than 10 kHz, the electrical conductivity starts increasing according to the power law.

Table 1 presents the comparison of DC electrical conductivity between cells with SE-1211, DODA CBDHA and DNA–DODA alignment layers. As can be seen, the electrical conductivity of the liquid crystal in all three types of cells is comparable and is of the order of $10^{-7}$ S m$^{-1}$. These are typical values for nematic liquid crystals. For example, classical liquid crystals based on cyanobiphenyls, such as 5CB and E7, exhibit an electrical conductivity of $10^{-7}$–$10^{-8}$ S m$^{-1}$.[32] Based on these results, it can be concluded that orienting layers containing DNA–DODA and DNA–CBDHA complexes do not affect the liquid crystal's ionic contaminations. Therefore, they can be successfully used in liquid crystal transducers.

## 4. Conclusions

The prepared and investigated DNA isolated from banana waste derivatives used as a homeotropic alignment layer of liquid crystal materials shows excellent properties in applications. Based on electro-optical measurements, it can be concluded that the alignment layers of DNA–DODA and DNA–CBDHA complexes are a good, cheap, and eco-friendly alternative to commercial synthetic polyimides. Therefore, they can be successfully used in liquid crystal transducers. Moreover, the proposed alignment films show more potential in biosensors and future green electronics.

## Author contributions

DW: writing – original draft, visualization, supervision, resources, methodology, investigation, formal analysis, data curation, and conceptualization. RW: writing – original draft, methodology, investigation, formal analysis, data curation, and conceptualization. AS: writing – original draft, methodology, investigation, formal analysis, data curation, and conceptualization. All authors have approved the final version of the manuscript.

## Data availability

The authors confirm that the data supporting the findings of this study are available within the article. The further derived data sets of this study are available from the corresponding author upon request.

## Conflicts of interest

There are no conflicts to declare.

## Acknowledgements

This work was supported by the National Centre of Science MINIATURA 6 2022/06/X/ST5/00508 and UGB 22-720 and 22-723. The authors thank Mateusz Mrukiewicz for performing ionic conductivity measurements.

## References

1 P. Sharma, V. K. Gaur, S.-H. Kim and A. Pandey, *Bioresour. Technol.*, 2020, **299**, 122580.
2 K. B. Arun, A. Madhavan, R. Sindhu, P. Binod, A. Pandey, R. Reshmy and R. Sirohi, *Ind. Crops Prod.*, 2020, **154**, 112621.
3 P. Sharma, V. K. Gaur, R. Sirohi, S. Varjani, S.-H. Kim and J. W. C. Wong, *Bioresour. Technol.*, 2021, **325**, 124684.

Table 1 The values of DC electrical conductivity $\sigma_{DC}$ and $a$ and $m$ parameters in the LC cells with SE-1211, DODA–CBDHA, and DNA–DODA alignment layers

| Alignment layer | Conductivity $\sigma$ [S m$^{-1}$] | $a$ | $m$ [F m$^{-1}$] |
| --- | --- | --- | --- |
| SE-1211 | $1.58 \times 10^{-7}$ | 1.33 | $1.39 \times 10^{-6}$ |
| DNA–DODA | $1.23 \times 10^{-7}$ | 1.41 | $1.55 \times 10^{-6}$ |
| DNA–CBDHA | $1.89 \times 10^{-7}$ | 1.35 | $1.23 \times 10^{-6}$ |

4 K. S. Ganesh, A. Sridhar and S. Vishali, Utilization of fruit and vegetable waste to produce value-added products: Conventional utilization and emerging opportunities-A review, *Chemosphere*, 2022, **287**(3), 132221.
5 FAO. 2022. FAOSTAT. https://www.fao.org/faostat/en/12.10.2023.
6 D. Salazar, M. Arancibia, D. Lalaleo, R. Rodríguez-Maecker, M. E. López-Caballero and M. P. Montero, *Food Hydrocolloids*, 2022, **122**, 107048.
7 C. Cecchini, *Des. J.*, 2017, **20**, S1596.
8 K. Gautam, R. Vishvakarma, P. Sharma, A. Singh, V. K. Gaur, S. Varjani and J. K. Srivastava, *Bioresour. Technol.*, 2022, **361**, 127650.
9 G. Sharmila, C. Muthukumaran, N. M. Kumar, V. M. Sivakumar and M. Thirumarimurugan, Chapter 12 - Food waste valorization for biopolymer production, *Curr. Dev. Biotechnol. Bioeng.: Resour. Recovery Wastes*, 2020, 233–249.
10 M. R. Kosseva, S. Zhong, M. Li, J. Zhang and N. A. Tjutju, Biopolymers produced from food wastes: A case study on biosynthesis of bacterial cellulose from fruit juices, *Food Industry Wastes: Assessment and Recuperation of Commodities*, 2020, pp. 225–254.
11 S. Xu, Y. Li, Y. Liu, J. Sun, H. Ren and S.-T. Wu, *Micromachines*, 2014, **5**, 300–324.
12 X. Zhan, Y. Liu, K.-L. Yang and D. Luo, *Biosensors*, 2022, **12**(8), 577.
13 M. Cieplak, R. Węgłowski, Z. Iskierko, D. Węgłowska, P. S. Sharma, K. R. Noworyta, F. D'souza and W. Kutner, *Sensors*, 2020, **20**(17), 4692.
14 M.-J. Lee and W. Lee, *Liq. Cryst.*, 2020, **47**(8), 1145–1153.
15 Z. Dang, B. Peng, D. Xie, S. Yao, M. Jiang and J. Bai, *Appl. Phys. Lett.*, 2008, **92**, 112910.
16 Z. Liu, F. Yu, Q. Zhang, Y. Zeng and Y. Wang, *Eur. Polym. J.*, 2008, **44**, 2718–2727.
17 X. Wang, P. Zhang, Y. Chen, L. Luo, Y. Pang and X. Liu, *Macromolecules*, 2011, **44**, 9731–9737.
18 R. D. Rusu and M. J. M. Abadie, in *New high-performance materials: bio-based, eco-friendly polyimides, Polyimides for electronic and electrical engineering applications*, ed. S. Diaham, IntechOpen, Londra, UK, 2020, ISBN: 978-1-83880-098-7.
19 I. Rau, J. G. Grote, F. Kajzar and A. Pawlicka, *C. R. Phys.*, 2012, **13**(8), 853.
20 A. Spadło, N. Bennis, R. Węgłowski, D. Węgłowska, K. Czupryński and J. M. Otón, *Mol. Cryst. Liq. Cryst.*, 2018, **657**, 56.
21 P. Marć, N. Bennis, A. Spadło, A. Kalbarczyk, R. Węgłowski, K. Garbat and L. R. Jaroszewicz, *Crystals*, 2019, **9**(8), 387.
22 J. G. Grote, D. E. Diggs, R. L. Nelson, J. S. Zetts, F. K. Hopkins, N. Ogata, J. A. Hagen, E. Heckman, P. P. Yaney, M. O. Stone and L. R. Dalton, *Mol. Cryst. Liq. Cryst.*, 2005, **426**(1), 3–17.
23 R. Węgłowski, W. Piecek, A. Kozanecka-Szmigiel, J. Konieczkowska and E. Schab-Balcerzak, *Opt. Mater.*, 2015, **49**, 224–229.
24 M. Mrukiewicz, P. Perkowski, M. Urbańska, D. Węgłowska and W. Piecek, *J. Mol. Liq.*, 2020, **317**(1), 113810.
25 Y.-W. Kwon, C. H. Lee, D.-H. Choi and J.-I. Jin, *J. Mater. Chem.*, 2009, **19**, 1353–1380.
26 W. C. Johnson Jr., CD of nucleic acids, in *Circular Dichroism: Principles and Applications*, ed. N. Berova, K. Nakanishi and R. W. Woody, Wiley-VCH, New York, 2nd edn, 2000, pp. 703–718.
27 A. Radko, S. Lalik, A. Deptuch, T. Jaworska-Gołąb, R. Ekiert, N. Górska, K. Makyla-Juzak, J. Nizioł and M. Marzec, *Polymer*, 2021, **235**, 124277.
28 B. L. Van Horn and H. H. Winter, *Appl. Opt.*, 2001, **40**, 2089–2094.
29 A. R. MacGregor, *J. Opt. Soc. Am. A*, 1990, **7**(3), 337–347.
30 A. K. Jonscher, *Nature*, 1977, **267**, 673–679.
31 M. R. Costa, R. A. C. Altafim and A. P. Mammana, *Liq. Cryst.*, 2001, **28**, 1779–1783.
32 S. Naemura and A. Sawada, *Mol. Cryst. Liquid Cryst.*, 2003, **400**, 79–96.